\documentclass{ws-ijmpd}
\usepackage{graphicx}
\usepackage{calligra}
\usepackage{amsmath}
\allowdisplaybreaks[1]
\DeclareMathAlphabet{\mathcalligra}{T1}{calligra}{m}{n}
\DeclareFontShape{T1}{calligra}{m}{n}{<->s*[2.2]callig15}{}
\newcommand{\be}{\begin{equation}}
\newcommand{\ee}{\end{equation}}
\newcommand{\ba}{\begin{align}}
\newcommand{\el}{\end{align}}
\newcommand{\bsp}{\begin{split}}
\newcommand{\esp}{\end{split}}

\newcommand{\mc}{\mathcalligra}

\begin{document}
\title{An improved method for constructing models of self-gravitating tori 
around black holes}

\author{Nikolaos Stergioulas}

\address{Department of Physics, Aristotle University of Thessaloniki \\
  Thessaloniki 54124, Greece}

\maketitle

\begin{history}
\received{Day Month Year}
\revised{Day Month Year}
\comby{Managing Editor}
\end{history}

\begin{abstract}
General-relativistic models of self-gravitating tori around black holes 
are constructed with a self-consistent-field method in compactified coordinates.
The numerical code is highly accurate and robust, allowing for the 
construction of models that exactly fill their Roche lobe, when a cusp exists. 
As a first application, we focus on self-consistent models with cusp, having 
different values of constant specific angular momentum. Scaling all results with 
the mass of the black hole, we find evidence that models with constant specific 
angular momentum that can fill their Roche lobe are still limited by $l<4M_{\rm BH}$ 
(as is the case for models constructed in a fixed background metric) even for heavy 
tori.
\end{abstract}

%%%%%%%%%%%%%%%%%%%%%%%%%%%%%%%%%%%%%%%%%%%%%%%%%%%%%%%%%%%%%%%
%%%%%%%%%%%%%%%%%%%%%%%%%%%%%%%%%%%%%%%%%%%%%%%%%%%%%%%%%%%%%%%
\section{Introduction}
\label{sec:Intro}
%%%%%%%%%%%%%%%%%%%%%%%%%%%%%%%%%%%%%%%%%%%%%%%%%%%%%%%%%%%%%%%
%%%%%%%%%%%%%%%%%%%%%%%%%%%%%%%%%%%%%%%%%%%%%%%%%%%%%%%%%%%%%%%

Massive relativistic accretion tori around black holes (BHs) can
form as transient structures in several astrophysical scenarios,
including the core-collapse of massive stars\cite{Woosley93b,Woosley06} and
the merger of neutron star (NS) and NS-BH binaries\cite{Ruffert96b,Duez10a,Rezzolla10b}. 
Recent numerical simulations have demonstrated that the mass of the torus 
resulting from binary NS-NS or BH-NS mergers, which are  
candidates for the central engine of short gamma-ray bursts (GRBs), 
can be in the range of $\sim 0.01 - 0.2M_{\odot}$
~\cite{ShibataTaniguchi06,Shibata09,Duez10a,Rezzolla10b}.
In addition, studies of the stability of accretion tori have revealed
that they can be subject to several types of
non-axisymmetric instabilities in a number of
circumstances\cite{Papaloizou84,Papaloizou85,Kojima86,Woodward94}, 
which can lead to highly variable and unstable accretion rates.  
Nonaxisymmetric instabilities have been recently shown to also exist in fully
general-relativistic models of massive accretion tori\cite{Korobkin}. 

Another important instability, the axisymmetric runaway 
instability\cite{Abramowicz83} in models that fill 
their Roche lobe, has been investigated extensively. 
Numerical simulations in a time-varying Schwarzschild background 
showed that the instability is active for tori
with constant specific angular momentum\cite{Font02a}.
For the same class of tori, the adoption of a time-varying Kerr background resulted in 
longer timescales for the instability to grow, as the rotation of the black hole 
increased\cite{Daigne2004}.  
Adopting a power-law distribution of the specific angular momentum 
with (even small) positive slope stabilizes the 
torus against the runaway 
instability\cite{Daigne1997,Lu2000,Daigne2004,Zanotti2005}. 
The instability was shown to be active for constant specific angular
momentum tori, even when taking self-gravity into account\cite{Nishida1996a,Nishida1996b}.
For self-gravitating, non-constant specific angular momentum tori, the presence 
of the instability still depends on the slope of the specific angular momentum law, with 
the range shrinking as the black hole rotation increases\cite{Abramowicz1998}. 
When self-gravity is taken into account and the torus loses more than a few
percent of its mass due to mass transfer, it was found that the instability
is active for different specific angular momentum laws\cite{Masuda1997,Masuda1998}.
However, more recent, fully general-relativistic simulations of some models of self-gravitating 
tori with either constant or non-constant specific angular momentum did not show 
signs of the instability on a dynamical timescale\cite{Montero2010}. 
The parameter space to be investigated for the conditions of the onset of the runaway 
instability is still large and there is a need for 
highly accurate self-consistent models, suitable as numerical initial data for
simulation codes. 

Analytic models of tori in a fixed spacetime were presented in
Refs. \refcite{AJS,Kozlowski1978}. These can serve as an initial guess for a 
self-consisted-field approach. Indeed, some first self-consistent models in 
quasi-isotropic coordinates were presented in Ref. \refcite{NE}, but with a 
numerical grid truncated at a finite
outer radius. A different numerical method was used for constructing self-consistent
models of uniformly rotating rings around black holes\cite{Ansorg2005}. More recently, 
a new numerical method was presented for constructing initial data of self-consistent models that 
are suitable for simulations using the moving puncture method\cite{Shi}. 
Since Roche-lobe filling models can have very large outer radii,
compared to the radius of the central black hole and, in addition, accurate gravitational
wave extraction in numerical simulations requires the outer numerical boundaries
to be placed at even larger distances from the source, the task of computing 
accurate initial data on such different length scales is challenging, but can be
met with a compactified grid. In particular, a global compactified grid can
be constructed with a re-definition of the radial coordinate, as was done 
in Ref. \refcite{Cook1994} for models of rotating neutron stars. This type of 
compactification is adopted here in order to remove the limitations of the method 
presented in Ref. \refcite{NE}.

The effect of self-gravity on the properties of a black-hole torus spacetime (and 
on the properties of the torus itself) will, of course, directly depend on the
ratio of torus mass to black hole mass. Here, we present evidence that even for
heavy tori, models with constant specific angular momentum, $l$, can
fill their Roche lobe only if it is limited by the same relation that holds 
for models constructed in a fixed background metric, $l<4M_{\rm BH}$, where 
$M_{\rm BH}$ is the mass of the black hole.

The paper is organized as follows: In Sec. \ref{sec:station} we summarize the
main definitions and describe the numerical setup.
In Sec. \ref{sec:model} we
discuss in detail a representative model, while Sec. \ref{sec:sequence} 
presents the main properties of a sequence of models with constant specific
angular momentum. A brief discussion is found in Sec. \ref{sec:discussion}.

Throughout the paper, we assume a signature of $(-,+,+,+)$ and set $c=G=1$.

%%%%%%%%%%%%%%%%%%%%%%%%%%%%%%%%%%%%%%%%%%%%%%%%%%%%%%%%%%%%%%%
%%%%%%%%%%%%%%%%%%%%%%%%%%%%%%%%%%%%%%%%%%%%%%%%%%%%%%%%%%%%%%%
\section{Tori in stationary, axisymmetric spacetimes}
\label{sec:station}
%%%%%%%%%%%%%%%%%%%%%%%%%%%%%%%%%%%%%%%%%%%%%%%%%%%%%%%%%%%%%%%
%%%%%%%%%%%%%%%%%%%%%%%%%%%%%%%%%%%%%%%%%%%%%%%%%%%%%%%%%%%%%%%

The details of our formalism and numerical method can be found in Ref. \refcite{Stergioulas2011}.
Here we only summarize the main equations:

We assume that the spacetime is stationary and axisymmetric, with circular velocity field, 
while the matter is described as a perfect fluid with total energy density $\epsilon$, 
rest mass density $\rho$ and pressure $p$. We further assume that 
the torus has a constant specific angular momentum $l:= -u_\phi/u_t$, where $u^\alpha$ is
the 4-velocity. 
The hydrostationary equilibrium equation can be written in the  
form
\be
     \frac{\nabla p}{\epsilon+p}  
                              = -\nabla \ln (-u_t)  + \frac{\Omega \nabla l}{1-\Omega l},
                                                           \label{hydrostat3}
\ee
where $\Omega:=u^\phi/u^t$ is the angular velocity
and has a first integral only under specific conditions. For a barotropic equation 
of state of the form $p=p(\rho)$ one can define the log-enthalpy 
\be
     H(p):=\int_0^p \frac{dp'}{\epsilon(p')+p'},
\ee
with, $dH/dp=(\epsilon+p)^{-1}$. Then, the l.h.s. of (\ref{hydrostat3}) becomes
$(\epsilon+p)^{-1}\nabla p=\nabla H$ and if $\Omega=\Omega(l)$, we can write
\be
\nabla H = -\nabla W,
\ee
where $-\nabla W$ is the effective gravity, 
given by the gradient of the effective potential
\be
     W= \ln (-u_t) - \int^l \frac{\Omega}{1-\Omega l} dl+{\rm const.}
\label{W}
\ee
It follows that the surfaces of constant pressure 
coincide with the equipotential surfaces and at the location of maximum density, 
$\nabla H=\nabla W=0$ (the constant in (\ref{W}) can be set by requiring that 
$W$ vanishes at infinity). 

%%%%%%%%%%%%%%%%%%%%%%%%%%%%%%%%%%%%%%%%%%%%%%%%%%%%%%%%%%%%%%%
%%%%%%%%%%%%%%%%%%%%%%%%%%%%%%%%%%%%%%%%%%%%%%%%%%%%%%%%%%%%%%%
%\section{Constant specific angular momentum}
%\label{sec:constl}
%%%%%%%%%%%%%%%%%%%%%%%%%%%%%%%%%%%%%%%%%%%%%%%%%%%%%%%%%%%%%%%
%%%%%%%%%%%%%%%%%%%%%%%%%%%%%%%%%%%%%%%%%%%%%%%%%%%%%%%%%%%%%%%
For the case of constant specific angular momentum the first integral of (\ref{hydrostat3})
 becomes
\be
     H+\ln(-u_t) = {\rm const.}
\ee 
and for the homentropic, polytropic equation of state (EOS)
\begin{align}
     p =& K \rho^\Gamma, \label{p}\\
     \epsilon =& \rho + \frac{p}{\Gamma-1}, \label{eps}
\end{align}
where $K$ is the polytropic constant and $\Gamma$ is the polytropic exponent, 
one obtains 
\be
H=\ln[1+\Gamma/(\Gamma-1)K\rho^{\Gamma-1}],
\label{H}
\ee 
The density distribution is given algebraically, as an implicit
function of $\Omega$ (for given EOS, $l_0$, inner radius $r_{\rm in}$ and stationary, axisymmetric
spacetime):
\be 
     \rho  
             = \left \{ \frac{\Gamma-1}{K\Gamma}\left[ \frac{u_{t,{\rm in}}}{u_t}
                 -1 \right ] \right \}^{\frac{1}{\Gamma-1}}.
\label{rhodist}
\ee

For our self-consistent numerical scheme we use the analytic AJS solutions\cite{AJS} as 
an initial guess. In isotropic coordinates $(t,r,\theta,\phi)$, the metric of 
the Schwarzschild spacetime is
\be
ds^2 = -\left [ \frac{1-M/2r}{1+M/2r} \right]^2 dt^2 
+\left(1+\frac{M}{2r}\right )^4
(dr^2+r^2 d\theta^2+r^2 \sin^2\theta d \phi^2),
\ee
where the isotropic radial coordinate $r$ is related to the Schwarzschild radial
coordinate $\mc r$ by $r=1/2\left[ {\mc r} -M + \sqrt{ {\mc r}^2-2M{\mc r}} \right]$.
Requiring $W(r=\infty)=0$, the effective potential is
\be
W = \ln ({-u_t}).
\ee

A torus of finite size extends between ${  r}={ r}_{\rm in}$ and ${  r}={  r}_{\rm out}$ 
in the equatorial plane. At the density maximum, $\nabla_{ r} p =0$ and the fluid elements there
move as free test particles with Keplerian angular momentum
\be 
l_{\rm K}=\frac{ \sqrt{M} \left( 1+M/2r \right)^6 r^2}{\left( 1-M/2r\right)^2
         \left(M+M^2/4r+r \right)^{3/2}},
\ee
has a minimum at the marginally stable circular orbit for test particles, 
${  r}_{\rm ms}= 1/2(5+2\sqrt{6})\simeq 4.950$, which restricts 
${ r}_{\rm max} > {  r}_{\rm ms}\Rightarrow l>l_{\rm ms}=3.674M$.
The radius of the marginally bound orbit is the location 
where $W=0$ for $l=4M$, i.e.
$r_{\rm mb}\simeq 2.914$. The largest of three roots of $l=l_{\rm K}$ 
corresponds to  ${ r}={  r}_{\rm max}$, while the intermediate root, ${ r}={  r}_{\rm cusp}$, 
corresponds to the existence of a cusp, where again  $\nabla_{ r} p =0$. 
For a given EOS and black-hole mass $M$ and 
for a given value of $l$, there exists a one-parameter family of different finite-size tori 
with ${  r}_{\rm  cusp} < {  r}_{\rm in} <{  r}_{\rm max}$, which all have the same 
${  r}_{  max}$. For a chosen ${  r}_{\rm in}$ in this range, one obtains the density 
distribution from (\ref{rhodist}). 
The parameter space of possible equilibrium configurations is further limited by the 
existence of the marginally bound orbit, $r_{\rm mb}$. 
For $ l< l_{\rm mb}=4M$, the effective potential $W$ is
always negative in the equatorial plane, while for $l=l_{\rm mb}$, $W=0$ at ${  r}_{  mb}$ 
(an inflection point) and the location of the cusp is at 
${  r}_{\rm in}={ r}_{\rm mb}$.   For $l>l_{\rm mb}$ there exists a region in the equatorial plane 
with $W>0$.  Thus, for any $l>l_{\rm mb}$, only
tori without a cusp are possible, with the inner radius limited by the 
location of $W=0$. 

%%%%%%%%%%%%%%%%%%%%%%%%%%%%%%%%%%%%%%%%%%%%%%%%%%%%%%%%%%%%%%%
%%%%%%%%%%%%%%%%%%%%%%%%%%%%%%%%%%%%%%%%%%%%%%%%%%%%%%%%%%%%%%%
%\section{AJS tori in a Schwarzschild background}
%\label{sec:AJSScw}
%%%%%%%%%%%%%%%%%%%%%%%%%%%%%%%%%%%%%%%%%%%%%%%%%%%%%%%%%%%%%%%
%%%%%%%%%%%%%%%%%%%%%%%%%%%%%%%%%%%%%%%%%%%%%%%%%%%%%%%%%%%%%%%

%%%%%%%%%%%%%%%%%%%%%%%%%%%%%%%%%%%%%%%%%%%%%%%%%%%%%%%%%%%%%%%
%%%%%%%%%%%%%%%%%%%%%%%%%%%%%%%%%%%%%%%%%%%%%%%%%%%%%%%%%%%%%%%
%\section{Self-gravitating tori in quasi-isotropic coordinates}
%\label{sec:Self}
%%%%%%%%%%%%%%%%%%%%%%%%%%%%%%%%%%%%%%%%%%%%%%%%%%%%%%%%%%%%%%%
%%%%%%%%%%%%%%%%%%%%%%%%%%%%%%%%%%%%%%%%%%%%%%%%%%%%%%%%%%%%%%%

For self-gravitating tori, we use the metric in quasi-isotropic coordinates 
\be
ds^2 = - e^{2\nu} dt^2+e^{2\alpha}(dr^2+r^2 d\theta^2)
            +e^{2(\gamma-\nu)}r^2\sin^2\theta(d\phi-\omega dt)^2,
\label{quasi-iso}
\ee
where $\nu, \gamma, \alpha$ and $\omega$ are metric functions that
depend only on the coordinates $r$ and $\theta$ (see Ref. \refcite{Stergioulas2003}). 
Because $\gamma$ and
$\nu$ are divergent at the event horizon, we will use, instead, the
functions\cite{NE} $B := e^\gamma$ and $\lambda := e^\nu$, with boundary 
conditions on the event horizon $B =0$, $\lambda =0$ and $\omega =\omega_h$,
where $\omega_h$ is the constant angular velocity of the horizon. The horizon 
can be chosen to be a sphere of constant radius $r=h_0$. On the axis of rotation, 
the condition $\alpha=\gamma-\nu$ ensures local flatness.

In the angular direction we use the coordinate $\mu:=\cos\theta$ instead of $\theta$,
while in the radial direction 
we use a compactified, nondimensional coordinate $s$, defined through
\be
r:=r_e \frac{s}{1-s},
\ee
where $r_e$ is a suitably chosen radius (here we use $r_e=r_{\rm out}$, 
so that the outer edge of the torus in the equatorial plane corresponds to $s=0.5$). Through 
this compactification, the infinite domain $r \rightarrow [0,+\infty)$ is mapped onto the finite domain
$s \rightarrow [0,1]$.  The location of the horizon is at $s=s_0$, so that 
$h_0 = r_e s_0/(1-s_0)$. 
 
The field equations for the three metric functions $\lambda, B$ and $\omega$ are of
elliptic type
\begin{align}
\nabla^2 \lambda =& S_\lambda(s,\mu), \\
\left( \nabla^2+\frac{(1-s)^3}{r_e^2 s}\frac{\partial}{\partial_s}-
\frac{(1-s)^2}{r_e^2s^2}\mu\frac{\partial}{\partial \mu} \right) B =& S_B(s,\mu), \\
\left( \nabla^2+\frac{2(1-s)^3}{r_e^2 s}\frac{\partial}{\partial_s}-
\frac{2(1-s)^2}{r_e^2s^2}\mu\frac{\partial}{\partial \mu} \right) \omega =& S_\omega(s,\mu), 
\end{align}
where $\nabla^2$ is the flat-space Laplacian, and $S_\lambda$, $S_B$ and $S_\omega$ are
source terms.  
These elliptic-type equations are inverted using appropriate Green's functions, while the
equation for the metric function $\alpha$ is integrated as an ordinary differential equation.

%%%%%%%%%%%%%%%%%%%%%%%%%%%%%%%%%%%%%%%%%%%%%%%%%%%%%%%%%%%%%%%
%%%%%%%%%%%%%%%%%%%%%%%%%%%%%%%%%%%%%%%%%%%%%%%%%%%%%%%%%%%%%%%
%\section{Numerical scheme}
%\label{sec:num}
%%%%%%%%%%%%%%%%%%%%%%%%%%%%%%%%%%%%%%%%%%%%%%%%%%%%%%%%%%%%%%%
%%%%%%%%%%%%%%%%%%%%%%%%%%%%%%%%%%%%%%%%%%%%%%%%%%%%%%%%%%%%%%%

While we use nondimensional units of $c=G=M_{\rm BH}^{\rm AJS}=1$ in the numerical procedure, 
the solution can then be scaled to any desired black hole mass. Our grid is equidistant 
in $(s,\mu)$ with $2001\times1001$ grid points.  
For a chosen polytropic index $\Gamma$, 
the solution space of AJS models is three-dimensional: the parameter $l$ controls 
the rotation of the torus, $r_e/h_0$ controls its size and $K$ (or 
$\rho_{\rm max}$) its mass (through the variation of the equation 
of state).  
 
Our fixed-point-iteration method is described in detail in Ref. \refcite{Stergioulas2011}.
Here, we note that our numerical procedure yields directly a self-consistent model, starting 
from the AJS solution as an initial guess. In order to construct models that exactly 
fill their Roche lobe, we must be able to compute models that violate this limit, in 
order to approach it by bisection. For this purpose, we also construct models starting 
with an AJS solution that overfills the Roche lobe, i.e. with a value of $W_{\rm in}$ 
larger than $W_{\rm cusp}$, but truncated at
$r_{\rm cusp}$ (similar models were constructed in other works, in order to initiate mass
accretion onto the black hole). Exploring the allowed parameter range, we then find models 
that exactly fill their Roche lobe, up to a single grid point.

%%%%%%%%%%%%%%%%%%%%%%%%%%%%%%%%%%%%%%%%%%%%%%%%%%%%%%%%%%%%%%%
%%%%%%%%%%%%%%%%%%%%%%%%%%%%%%%%%%%%%%%%%%%%%%%%%%%%%%%%%%%%%%%
%\section{Derived properties}
%\label{sec:derived}
%%%%%%%%%%%%%%%%%%%%%%%%%%%%%%%%%%%%%%%%%%%%%%%%%%%%%%%%%%%%%%%
%%%%%%%%%%%%%%%%%%%%%%%%%%%%%%%%%%%%%%%%%%%%%%%%%%%%%%%%%%%%%%%

\begin{table}[th]
\tbl{Properties of a representative self-consistent model of a BH-torus system, 
with $N=3$, $l/M_{\rm BH}$=3.911, $r_{\rm out}/h_0$=49.005, $\omega_h=0$ and $K/M_{\rm BH}^{2/N}$=17.339. 
All radii are coordinate radii in the quasi-isotropic metric.}
{\begin{tabular}{@{}lll@{}} \toprule
%Property &   &    \\ \colrule
% 1.0  & \hphantom{0}281.0 & \hphantom{0}280.81 & 0.07 \\
% 0.1\hphantom{00} & \hphantom{0}876.0 & \hphantom{0}875.74 & 0.03 \\
% 0.01\hphantom{0} & 2441.0 & 2441.0\hphantom{0} & 0.0\hphantom{0} \\
% 0.001 & 4130.0 & 4129.3\hphantom{0} & 0.16\\ \botrule
Asymptotic spacetime mass        &       $M/M_{\rm BH} $ &    \hphantom{0}1.1730  \\
Torus mass                      &        $M_{\rm T}/M_{\rm BH}$   &  \hphantom{0}0.2097 \\
Maximum density                 & $\rho_{\rm max}M_{\rm BH}^2$  &   \hphantom{0}6.6126$\times 10^{-5}$ \\
Inner torus radius               &     $r_{\rm in}/M_{\rm BH}$  &   \hphantom{0}3.0413 \\
Radius of maximum density         &   $r_{\rm max}/M_{\rm BH}$   &  \hphantom{0}8.1931 \\
Outer torus radius                &     $r_{\rm out}/M_{\rm BH}$   &   23.6400 \\
Torus rest mass                   &      $M_0/M_{\rm BH}$  &   \hphantom{0}0.1964 \\
Torus internal energy              &     $U_{\rm T}/M_{\rm BH}$   &  \hphantom{0}2.9878$\times 10^{-3}$ \\
Torus rotational energy            &     $T_{\rm T}/M_{\rm BH}$  &   \hphantom{0}9.5559$\times 10^{-3}$ \\
Torus gravitational potential energy &   $ W_{\rm T}/M_{\rm BH}$ &   -3.5865$\times 10^{-2}$  \\
Torus rotational to potential energy &   $|T/W|_{\rm T}$    &  \hphantom{0}0.2664 \\
Torus angular momentum               &  $J_{\rm T}/M_{\rm BH}^2$  &  \hphantom{0}0.7390 \\ 
Black hole horizon radius                   &       $h_0/M_{\rm BH}$  & \hphantom{0}0.4824 \\
Black hole Komar charge           &      $M_{\rm H}/M_{\rm BH}$   &  \hphantom{0}0.9633 \\ \botrule
\end{tabular} \label{tab:model}}
\end{table}

%%%%%%%%%%%%%%%%%%%%%%%%%%%%%%%%%%%%%%%%%%%%%%%%%%%%%%%%%%%%%%%
%%%%%%%%%%%%%%%%%%%%%%%%%%%%%%%%%%%%%%%%%%%%%%%%%%%%%%%%%%%%%%%
\section{A representative model}
\label{sec:model}
%%%%%%%%%%%%%%%%%%%%%%%%%%%%%%%%%%%%%%%%%%%%%%%%%%%%%%%%%%%%%%%
%%%%%%%%%%%%%%%%%%%%%%%%%%%%%%%%%%%%%%%%%%%%%%%%%%%%%%%%%%%%%%%

\begin{figure}
  \centering 
  \resizebox{1.0\textwidth}{!}{
   %\hspace{-1.8 cm} 
   \includegraphics*{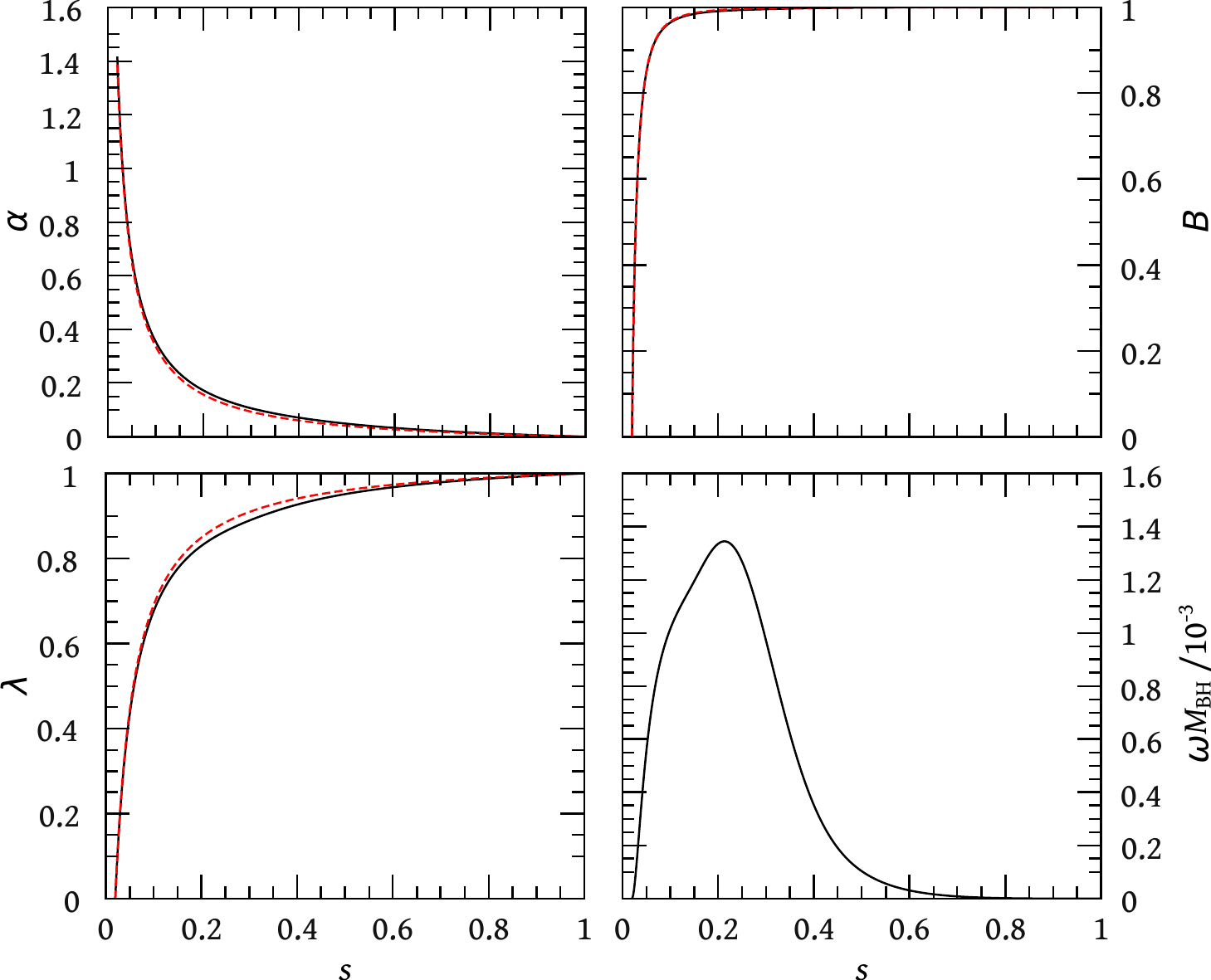}
   %\hspace{0.2 cm}
  }
  \caption{Distribution of the four metric functions of the representative model
of Table \ref{tab:model} in the equatorial plane (solid lines), as a function of the 
nondimensional compactified coordinate $s$ (the outer edge of the torus
is at $s=0.5)$. The horizon of the black hole is at $s_0=0.02$. 
The corresponding Schwarzschild solution (the spacetime used
in the AJS limit) is shown as dashed lines.}
  \label{fig:1}
\end{figure}

\begin{figure}
  \centering 
  \resizebox{1.0\textwidth}{!}{
   %\hspace{-1.8 cm} 
   \includegraphics*{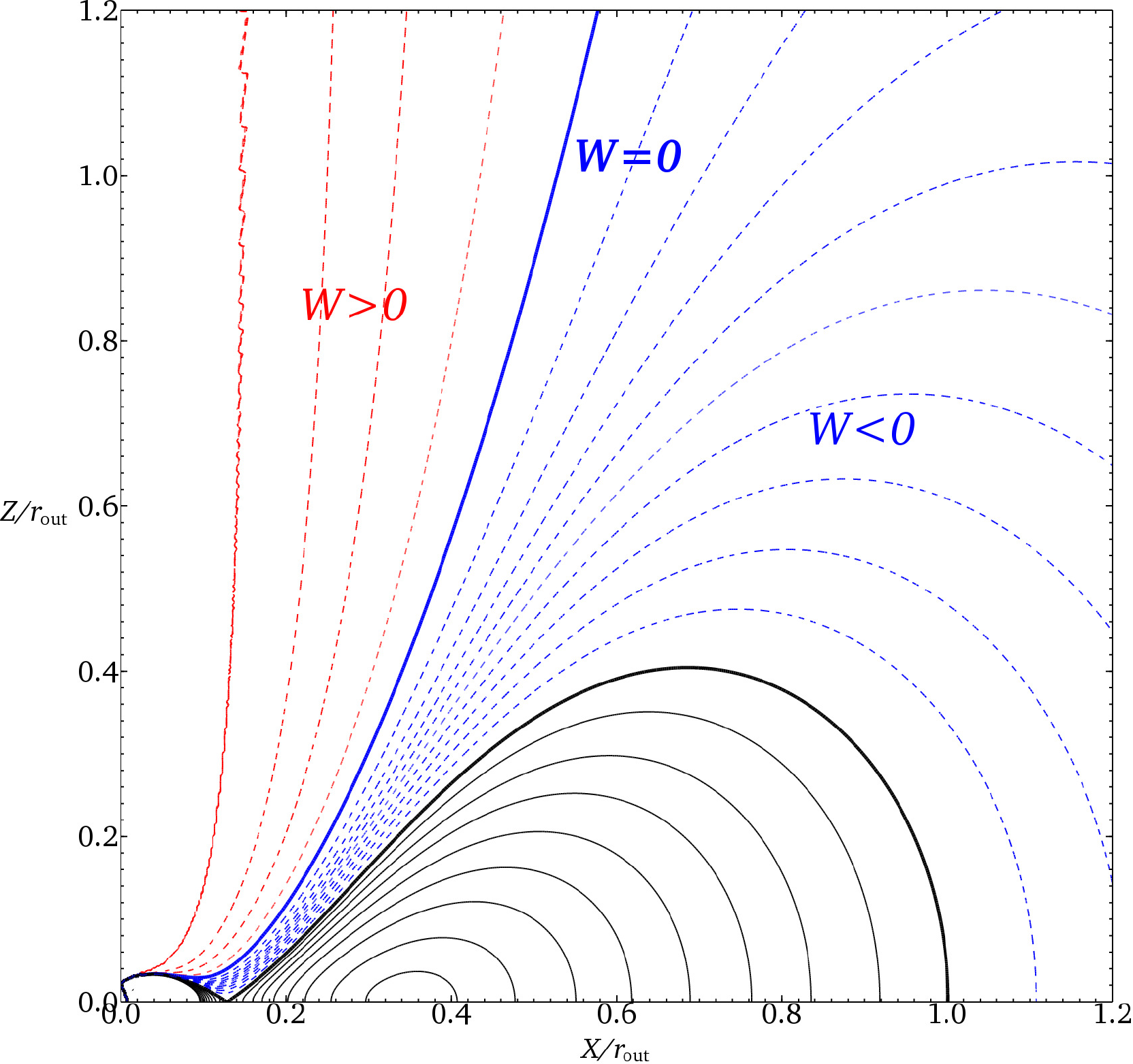}
   %\hspace{0.2 cm}
  }
  \caption{Isocontours of the effective potential $W$, in Cartesian coordinates, scaled
by the outer radius of the torus, $r_{\rm out}$, for the representative model
of Table \ref{tab:model}. The torus fills exactly its Roche lobe 
(thick black line), which has a cusp in the equatorial plane. Inside the Roche lobe, 
the isocontours of $W$ and of the density, $\rho$, coincide (thin black lines). Isocontours of
$W$ outside the torus can have negative or positive values (dashed lines), separated
by the marginally-bound surface ($W=0$). The horizon of the black hole is a sphere with 
radius $h_0/r_{\rm out}=0.02$ (not shown here).}
  \label{fig:2}
\end{figure}

\begin{figure}
  \centering 
  \resizebox{1.0\textwidth}{!}{
   %\hspace{-1.8 cm} 
   \includegraphics*{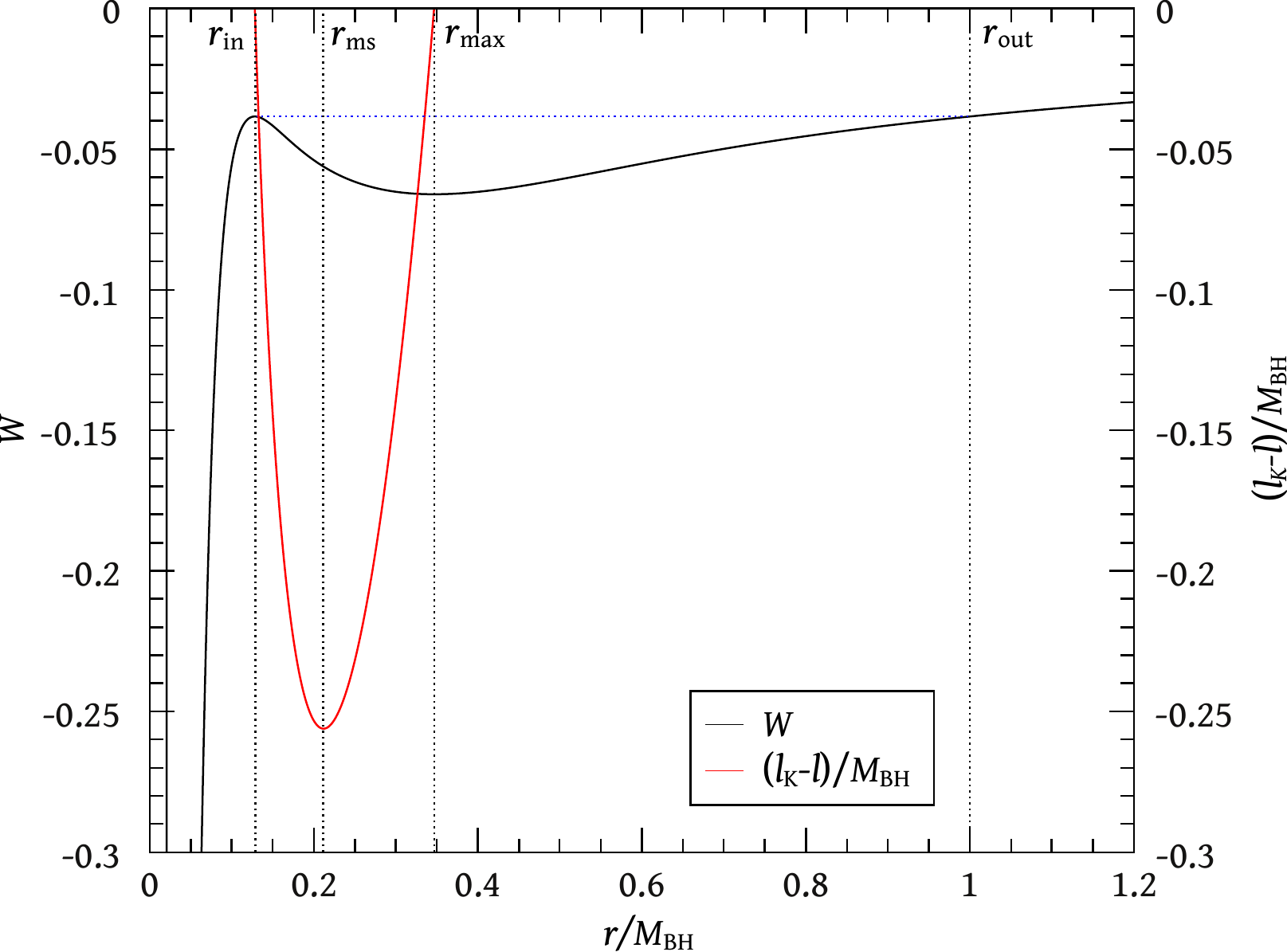}
   %\hspace{0.2 cm}
  }
  \caption{The effective potential $W$ and $l_{\rm K}-l$ in the equatorial plane
for the representative model of Table \ref{tab:model}. The radius of the inner edge of
the torus, $r_{\rm in}$, the marginally stable orbit $r_{\rm ms}$, the location
of the density maximum, $r_{\rm max}$, and the radius of the outer edge of
the torus, $r_{\rm out}$, are shown with dotted vertical lines. A dotted horizontal 
line connects the inner and outer edges of the torus at the same value of the
effective potential.}
  \label{fig:3}
\end{figure}

A representative solution is constructed for a polytropic
 equation of state with index $N=3$ and constant $K/M_{\rm BH}^{2/N}$=17.34, 
a ratio of outer torus radius to horizon radius $r_e/h_0$=49 and with $\omega_h=0$. 
The specific angular momentum in the initial guess is $l/M^{\rm AJS}_{\rm BH}$=3.675, 
while it is $l/M_{\rm BH}$=3.911 in the converged, self-consistent
solution. A detailed list of various properties, scaled by the mass of
the black hole, is displayed in Table 
\ref{tab:model} (see  Ref. \refcite{Stergioulas2011} for definitions). 
The ratio of torus mass to black hole mass is $M_{\rm T}/M_{\rm BH}$=0.21, 
while the total mass of the spacetime is $M/M_{\rm BH}$=1.17. 
In these coordinates, the horizon of the black hole is at a radius of 
$h_0/M_{\rm BH}$=0.482 (compared to 0.5 for the AJS solution). 
The torus itself extends from $r_{\rm in}/M_{\rm BH}$=3.04 to $r_{\rm out}/M_{\rm BH}$=23.64.
The relation $M_{\rm H}/M_{\rm BH} < (M-M_0)/M_{\rm BH} < 1$ holds
as expected\cite{Shi}. 

Notice that, the gravitational potential energy of the torus is
\be
W_{\rm T} := M-M_{\rm BH} -M_0-T_{\rm T}-U_{\rm T},
\ee
and includes both self-energy and binding energy between the
black hole and the torus. While all previous quantities are determined with
high accuracy (as was checked by proper convergence studies with grids of different sizes), 
this is not the case for $W_{\rm T}$, since it results as the difference
between considerably larger quantities. Even for the largest-size grid we can afford at
present, a convergence study showed that the numerical error in $W_{\rm T}$ can range 
from a few \% up to roughly 10\%, depending on the mass of the torus. This affects also
the ratio $|T/W|_{\rm T}$. We will thus report these two quantities only up to such an
uncertainty.

The four panels in Fig. \ref{fig:1} show the distribution of the four metric functions
$\alpha$, $B$, $\lambda$ and $\omega$, as a function of the nondimensional compactified
coordinate $s$ in the equatorial plane (we remind here that the outer radius of the 
torus is at $s=0.5$). The 
first three functions (solid lines) are compared to the corresponding Schwarzschild solution.
The largest impact of the self-gravity of the torus is in the metric function $\lambda$, 
while $B$ remains largely unaffected. The metric function $\omega$ vanishes for the 
Schwarzschild metric, while in the self-consistent solution it shows a peak inside the
torus. At the horizon and at infinity, all metric functions attain their expected
boundary values, enforced by the Green's functions approach. 

Fig. \ref{fig:2} shows the isocontours of the effective potential $W$ in Cartesian 
coordinates $(X,Z)$ scaled by the outer radius of the torus. For this particular model  
the marginally bound isocontour, $W=0$, does not cross the equatorial plane. Instead, all 
isocontours that cross the equatorial plane have $W<0$ and a cusp exists. Matter fills
exactly the Roche lobe (thick black line), which has a cusp in the equatorial plane. 
Inside the Roche lobe, the isocontours of the effective potential $W$ coincide with 
corresponding isodensity contours.

Fig. \ref{fig:3} shows the distribution of the effective potential $W$ in the equatorial
plane, as well as the difference of the Keplerian specific angular momentum $l_{\rm K}$ 
from the constant value $l$ inside the torus. For this model with cusp, the inner 
edge of the torus coincides with a local maximum in $W$ and with the location where 
$l_{\rm K}-l=0$. The close agreement between these two quantities demonstrates the 
high accuracy of the numerical solution. The radius of the marginally stable orbit, 
$r_{\rm ms}$, corresponds to the local minimum in $l_{\rm K}-l$, while the radius where
the density has a maximum, $r_{\rm max}$, coincides with a local minimum in $W$. 
Furthermore, at the outer radius of the torus, $r_{\rm out}$, the effective potential
has the same value as at the cusp.

%%%%%%%%%%%%%%%%%%%%%%%%%%%%%%%%%%%%%%%%%%%%%%%%%%%%%%%%%%%%%%%
%%%%%%%%%%%%%%%%%%%%%%%%%%%%%%%%%%%%%%%%%%%%%%%%%%%%%%%%%%%%%%%
\section{A sequence of models with cusp}
\label{sec:sequence}
%%%%%%%%%%%%%%%%%%%%%%%%%%%%%%%%%%%%%%%%%%%%%%%%%%%%%%%%%%%%%%%
%%%%%%%%%%%%%%%%%%%%%%%%%%%%%%%%%%%%%%%%%%%%%%%%%%%%%%%%%%%%%%%

\begin{table}[h]
\tbl{Properties of a sequence of models with a cusp 
with $N=3$, $l/M^{\rm AJS}_{\rm BH}$=3.675 and $\omega_h=0$.}
{\begin{tabular}{@{}ccccccc@{}} \toprule
$M_{\rm T}/M_{\rm BH}$  & $M/M_{\rm BH} $ & $l/M_{\rm BH}$  &  $r_{\rm in}/M_{\rm BH}$ 
&  $r_{\rm max}/M_{\rm BH}$ & 
 $r_{\rm out}/M_{\rm BH}$ &$J_{\rm T}/M_{\rm BH}^2$ \\ \colrule
--    & 1.000 &  3.675   &  4.838  & 5.092 &  \hphantom{00}5.244   & --       \\
0.134 & 1.107 &  3.859   &  3.193  & 7.763 &  \hphantom{0}18.235   & 0.460    \\
0.296 & 1.249 &  3.948   &  2.888  & 8.765 &  \hphantom{0}39.329  & 1.079    \\
0.366 & 1.312 &  3.952   &  2.784  & 9.025 &  \hphantom{0}58.745  & 1.357    \\
0.438 & 1.379 &  3.947   &  2.723  & 9.167 &  \hphantom{0}85.049  & 1.644    \\
0.546 & 1.484 &  3.948   &  2.660  & 9.293 &  133.360 & 2.074    \\\botrule
\end{tabular} \label{tab:sequence}}
\end{table}

\begin{figure}
  \centering 
  \resizebox{1.0\textwidth}{!}{
   %\hspace{-1.8 cm} 
   \includegraphics*{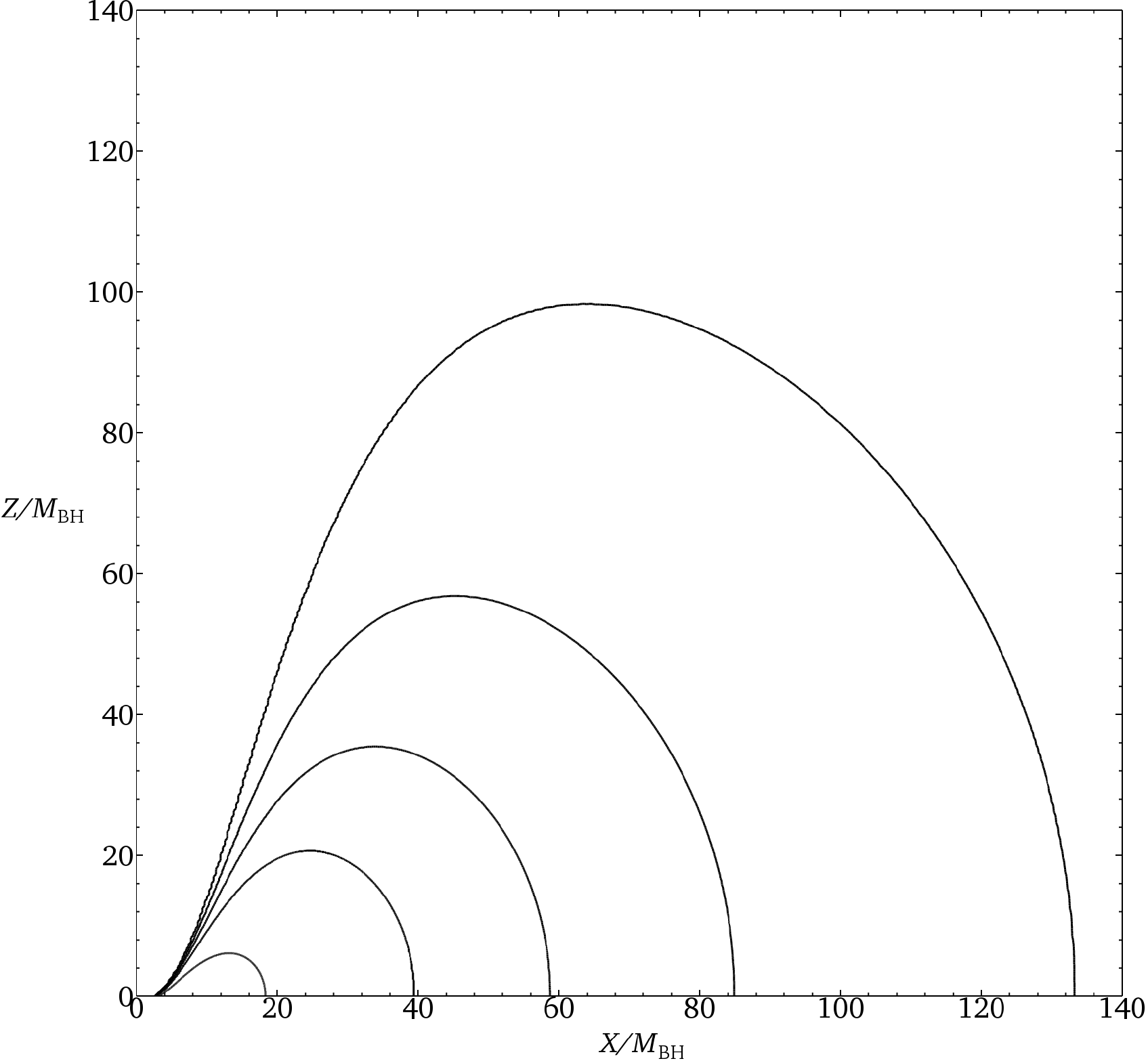}
   %\hspace{0.2 cm}
  }
  \caption{Outer surface for each of the models presented in Table \ref{tab:sequence} (except 
for the AJS limit) in Cartesian coordinates. With increasing torus mass, the outer 
edge of the torus, as well as its vertical extent, increase significantly. The horizon of 
the black hole in each case is a sphere with radius ranging from $h_0/M_{\rm BH}=0.47$ (largest 
torus) to $h_0/M_{\rm BH}=0.49$ (smallest torus) and is not shown here.}
  \label{fig:4}
\end{figure}

\begin{figure}
  \centering 
  \resizebox{1.0\textwidth}{!}{
   %\hspace{-1.8 cm} 
   \includegraphics*{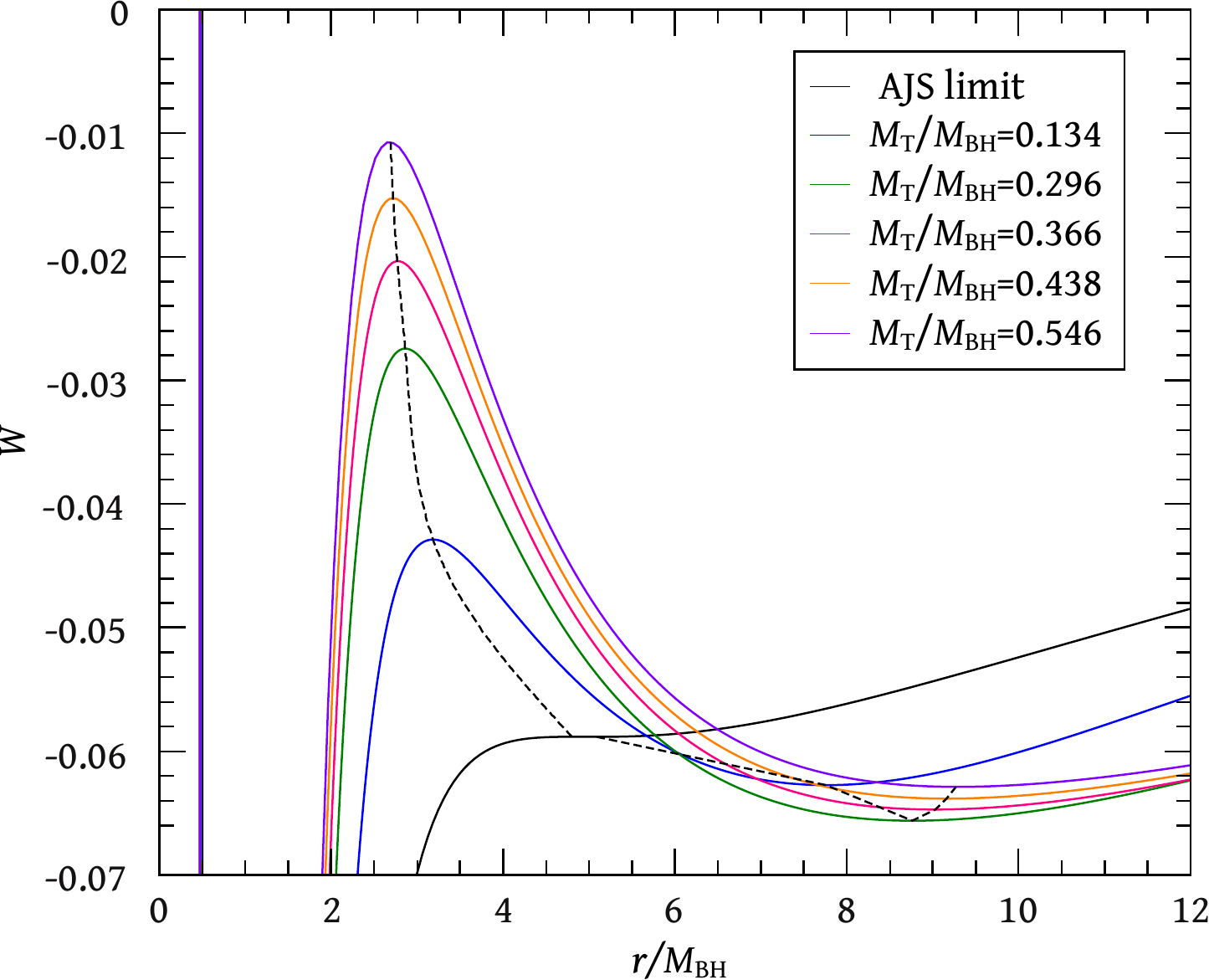}
   %\hspace{0.2 cm}
  }
  \caption{The effective potential in the equatorial plane, for the sequence of models 
presented in Table \ref{tab:sequence}. The shape of the potential changes significantly
with increasing torus mass, due to self-gravity effects. A dashed line connects the
local maximum for each model, that corresponds to $r_{\rm in}$, while another dashed line 
connects the local minimum for each model, that corresponds to $r_{\rm max}$.}
  \label{fig:5}
\end{figure}

\begin{figure}
  \centering 
  \resizebox{1.0\textwidth}{!}{
   %\hspace{-1.8 cm} 
   \includegraphics*{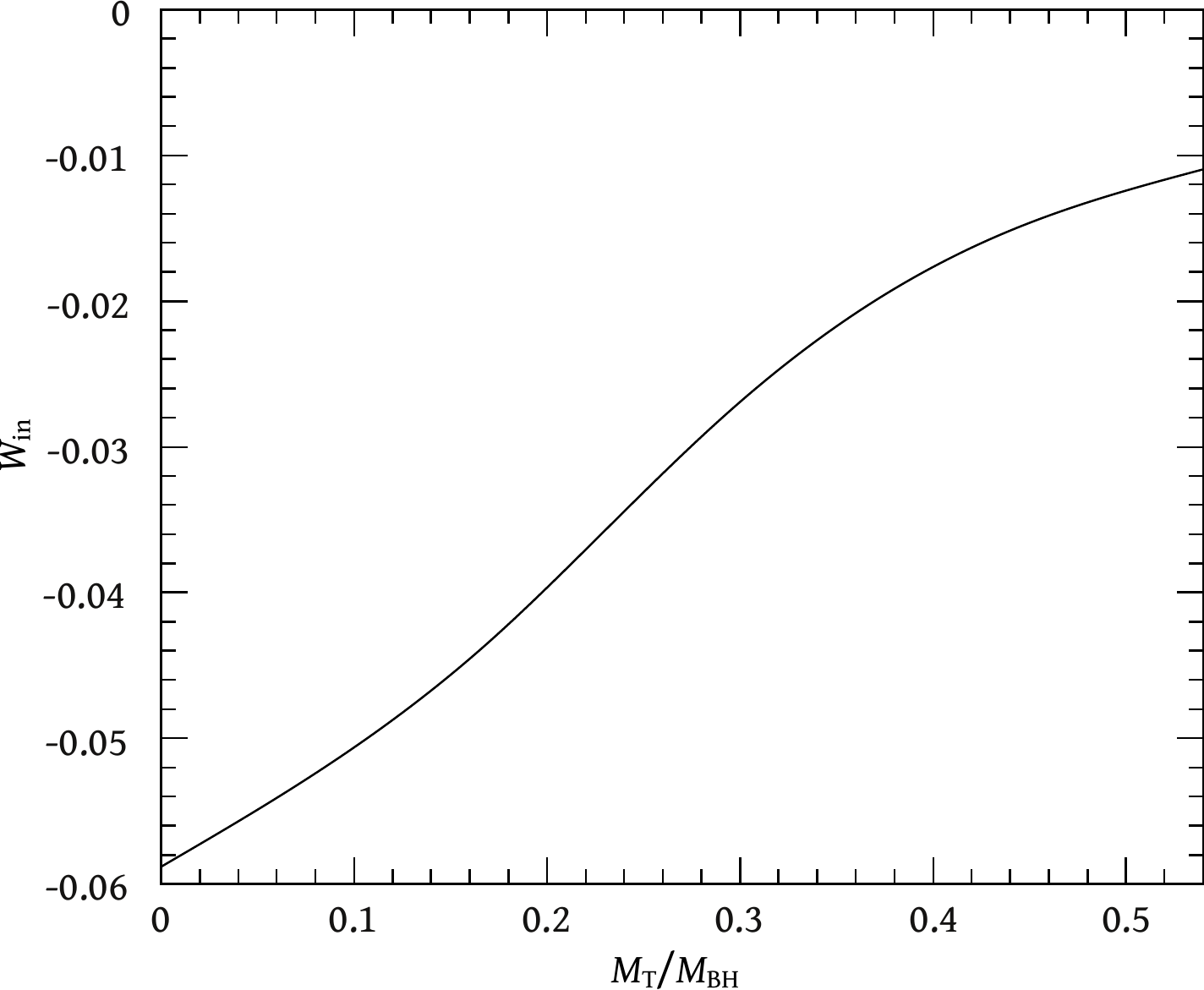}
   %\hspace{0.2 cm}
  }
  \caption{The value of the effective potential $W_{\rm in}$ at the inner edge of the 
torus for the sequence of models presented in Table \ref{tab:sequence}. With increasing
torus mass, $W_{\rm in}$ approaches, but does not cross the marginally-bound value of 
$W_{\rm in}=0$.}
  \label{fig:6}
\end{figure}

In the AJS limit, models with constant specific angular momentum
are required to have $l>3.674M^{\rm AJS}_{\rm BH}$, in order for a 
finite-size torus to exist. We construct a sequence of self-consistent 
models with different torus mass, that all fill exactly their Roche lobe,
with $l>3.675M^{\rm AJS}_{\rm BH}$. Table \ref{tab:sequence} summarizes
the main properties of models belonging to this sequence, with the
first line corresponding to the AJS solution (no values for $M_{\rm T}$ 
and $J_{\rm T}/M_{\rm BH}^2$ are given, since the AJS solution is
independent of the torus mass). In the AJS limit, the torus has
a very small size, extending between $r_{\rm in}/M_{\rm BH}=4.84$ 
and  $r_{\rm out}/M_{\rm BH}=5.24$. In the self-consistent solution, 
the value of $l/M_{\rm BH}$ changes, but in the last four models
of the sequence it stays practically constant at $l/M_{\rm BH}=3.95$.
Increasing the mass of the torus only increases the total mass-energy
of the spacetime, but not the mass of the BH. Along the sequence, 
the radius of the inner edge of the torus decreases, approaching the
horizon, while the radius of the maximum density increases. The
most dramatic increase is in the outer radius of the torus, which 
grows more than 25 times for the most massive model considered here,
compared to the AJS limit. Part of this difference can be attributed
to the fact that $l/M_{\rm BH}$ increases. Nevertheless, the last
four models in Table \ref{tab:sequence} constitute a proper sequence
of models with practically the same $l/M_{\rm BH}$, so they can
be compared on an equal footing. Among these models, the outer radius
also increases significantly with increasing mass of the torus. 
This can been seen in more detail in Fig. \ref{fig:4}, which 
compares the structure of the different tori in Table \ref{tab:sequence}
(showing only the surface for each model), in Cartesian coordinates
scaled by $M_{\rm BH}$. The total vertical extent of the torus 
approaches twice its extent in the equatorial plane, as the torus
mass increases.

Fig. \ref{fig:5} shows the effective potential in the equatorial plane, 
for the sequence of models in Table \ref{tab:sequence}. As the torus
mass increases, the value of the effective potential at the inner edge 
of the torus, $W_{\rm in}$, (shown with a dashed line connecting the local
maxima in the effective potential) increases significantly, which can be seen
in more detail in Fig. \ref{fig:6}. It appears that $W_{\rm in}$ only
asymptotically approaches the marginally-bound value of $W=0$ and does not
cross this limit. This is natural, since marginally bound tori close
their Roche lobe at infinity. Thus, a sequence of bound models with
increasing mass would first need to reach a marginally bound state with
infinite radius, before connecting to a sequence of models that have no
cusp. We conclude that for tori with constant specific angular momentum, 
the limitation of $l<4M_{\rm BH}$ may also hold for self-consistent models, 
as it does in the AJS limit, although a more detailed investigation is needed
to locate the precise limit.

%%%%%%%%%%%%%%%%%%%%%%%%%%%%%%%%%%%%%%%%%%%%%%%%%%%%%%%%%%%%%%%
%%%%%%%%%%%%%%%%%%%%%%%%%%%%%%%%%%%%%%%%%%%%%%%%%%%%%%%%%%%%%%%
\section{Discussion}
\label{sec:discussion}
%%%%%%%%%%%%%%%%%%%%%%%%%%%%%%%%%%%%%%%%%%%%%%%%%%%%%%%%%%%%%%%
%%%%%%%%%%%%%%%%%%%%%%%%%%%%%%%%%%%%%%%%%%%%%%%%%%%%%%%%%%%%%%%

We present an improved method for constructing general-relativistic 
models of self-gravitating tori around black holes. It is a 
self-consistent-field method, with the field equations
inverted using Green's functions and its improvement over an 
existing approach\cite{NE} consists in the use of compactified coordinates. This allows for the metric
to be computed up to spatial infinity, resulting in highly accurate initial
data for numerical evolution codes. These initial data have already
been used successfully in studying nonaxisymmetric instabilities  
in fully general-relativistic models of massive accretion 
tori\cite{Korobkin}. Of interest is also a more detailed 
study of the precise conditions under which the axisymmetric
runaway instability is active, since the numerical code is highly 
accurate and robust, allowing for the construction of models that 
exactly fill their Roche lobe, when a cusp exists. 

Scaling all results with the mass of the black hole, we found 
evidence that models with constant specific angular momentum that can
fill their Roche lobe are still limited by $l<4M_{\rm BH}$ (as is the case
for models constructed in a fixed background metric) even for heavy tori.
The precise limit must be determined with a more detailed study. 
More generally, we plan to investigate the allowed parameter space
for different types of self-gravitating tori, including the case of 
rotating black holes and more general rotation laws and equations of state.

\section*{Acknowledgments}
I would like to thank Burkhard Zink, Oleg Korobkin and Marcus Ansorg for useful discussions 
and I'm grateful to the University of Tuebingen for hospitality during an extended visit.


\begin{thebibliography}{9}
\bibitem{Woosley93b} Woosley S E 1993 {\it Astrophys. J.} {\bf 405} 273
\bibitem{Woosley06} Woosley S E and Bloom J S 2006 {\it Annual Rev. Astron. Astrophys.} {\bf 44} 507
\bibitem{Ruffert96b} Ruffert M, Janka H Th and Sch{\"a}fer G 1996 {\it Astron. Astrophys.} {\bf 311} 532
\bibitem{Duez10a} Duez M D 2010 {\it Class. Quant. Grav.} {\bf 27} 114002
\bibitem{Rezzolla10b} Rezzolla L, Baiotti L, Giacomazzo B, Link D and Font J A 2010 
      {\it Class. Quant. Grav.} {\bf 27} 114105
\bibitem{ShibataTaniguchi06} Shibata M and Taniguchi K 2006 {\it Phys. Rev. D} {\bf 73} 064027
\bibitem{Shibata09} Shibata M, Kyutoku K, Yamamoto T and Taniguchi K 2009 {\it Phys. Rev. D} {\bf 79} 044030
\bibitem{Papaloizou84} Papaloizou J C B and Pringle J E 1984 {\it Mon. Not. R. Astron. Soc.} {\bf 208} 721
\bibitem{Papaloizou85} Papaloizou J C B and Pringle J E 1985 {\it Mon. Not. R. Astron. Soc.} {\bf 213} 799
\bibitem{Kojima86} Kojima Y 1986 {\it Prog. Theor. Phys.} {\bf 75} 251
\bibitem{Woodward94} Woodward J W, Tohline J E and Hachisu I 1994 {\bf 420} 247
\bibitem{Korobkin} Korobkin O, Abdikamalov E B, Schnetter E, Stergioulas N and Zink B 2010
 {\it Phys. Rev. D} {\bf 83} 043007
\bibitem{Abramowicz83} Abramowicz M A, Calvani M and Nobili L 1983 {\it Nature} {\bf 302} 597
\bibitem{Font02a} Font J A and Daigne F 2002 {\it Mon. Not. R. Astron. Soc.} {\bf 334} 383
\bibitem{Daigne2004} Daigne F and Font J A {\it Mon. Not. R. Astron. Soc.} {\bf 349} 841
\bibitem{Daigne1997} Daigne F and Mochkovitch R {\it Mon. Not. R. Astron. Soc.} {\bf 285} L15
\bibitem{Abramowicz1998} Abramowicz M A, Karas V and Lanza A 1998  {\it Astron. Astrophys.} {\bf 331} 1143
\bibitem{Lu2000} Lu Y, Cheng K S, Yang L T and Zhang L {\it Mon. Not. R. Astron. Soc.} {\bf 314} 453
\bibitem{Zanotti2005} Zanotti O, Font J A, Rezzolla L and Montero P 2005 
 {\it Mon. Not. R. Astron. Soc.} {\bf 356} 1371
\bibitem{Nishida1996a} Nishida S and Eriguchi Y 1996 {\it Astrophys. J.} {\bf 461} 320
\bibitem{Nishida1996b} Nishida S, Lanza A, Eriguchi Y and Abramowicz M A 1996 
{\it Mon. Not. R. Astron. Soc.} {\bf 278} L41
\bibitem{Masuda1997} Masuda N and Eriguchi Y 1997 {\it Astrophys. J.} {\bf 489} 804
\bibitem{Masuda1998} Masuda N, Nishida S and Eriguchi Y 1998 {\it Mon. Not. R. Astron. Soc.} {\bf 297} 1139
\bibitem{Montero2010} Montero P J, Font J A, Shibata M 2010 {\it Phys. Rev. Letters} 
       {\bf 104} 191101
\bibitem{AJS} Abramowicz M, Jaroszynski M and Sikora M 1978 {\it Astron. Astrophys.} {\bf 63} 221
\bibitem{Kozlowski1978} Kozlowski M, Jaroszy\'nski M and Abramowicz M A 1978 {\it Astron. Astrophys.}
 {\bf 63} 209
\bibitem{NE} Nishida S and Eriguchi Y 1994 {\it Astrophys. J.} {\bf 427} 429
\bibitem{Ansorg2005} Ansorg M and Petroff D 2005 {\it Phys. Rev. D} {\bf 72} 024019
\bibitem{Shi} Shibata M 2007 {\it Phys. Rev. D} {\bf 76} 064035
\bibitem{Cook1994} Cook G B, Shapiro S L and Teukolsky S A 1994 {\it Astrophys. J.} {\bf 422} 227
\bibitem{Stergioulas2011} Stergioulas N 2011 {\it J. Phys. Conf. Ser.} {\bf 283} 012036
\bibitem{Stergioulas2003} Stergioulas N 2003 {\it Living Rev. Relativity} {\bf 6} 3
\end{thebibliography}
\end{document}